\documentclass[prd,twocolumn,aps,preprintnumbers,showpacs]{revtex4}
\usepackage{amssymb}
\usepackage{epsfig}
\usepackage{graphicx}
\usepackage{dcolumn}
\usepackage{bm}
\begin{document}
\preprint{BNL-NT-05/22, RBRC-529}
\title{Next-to-leading order QCD corrections to single-inclusive hadron production
in transversely polarized $\mathbf{pp}$ and $\mathbf{\bar{p}p}$
collisions}
\author{A.\ Mukherjee}
\affiliation{Instituut-Lorentz, University of Leiden, 2300 RA Leiden, The Netherlands}
\author{M.\ Stratmann}
\affiliation{Institut f{\"u}r Theoretische Physik, Universit{\"a}t Regensburg,
D-93040 Regensburg, Germany}
\author{W.\ Vogelsang}
\affiliation{Physics Department and RIKEN-BNL Research Center, Brookhaven National Laboratory,
Upton, New York 11973, U.S.A.}
%
%
\begin{abstract}
\noindent We present a calculation of the next-to-leading order
QCD corrections to the partonic cross sections contributing to
single-inclusive high-$p_T$ hadron production in collisions of
transversely polarized hadrons. We use a recently developed
projection technique for treating the phase space integrals in the
presence of the $\cos(2\Phi)$ azimuthal-angular dependence
associated with transverse polarization. Our phenomenological
results show that the double-spin asymmetry
$A_{\mathrm{TT}}^{\pi}$ for neutral-pion production is expected to be very
small for polarized $pp$ scattering at RHIC and could be much
larger for the proposed experiments with an asymmetric $\bar{p}p$
collider at the GSI.

\end{abstract}

\pacs{12.38.Bx, 13.85.Ni, 13.88.+e}
\maketitle

\section{Introduction}
%
\noindent
The partonic structure of spin-1/2 targets at the leading-twist level
is characterized entirely by the unpolarized, longitudinally polarized,
and transversely polarized distribution functions $f$, $\Delta f$,
and $\delta f$, respectively \cite{ref:jaffeji}.
Of these, the ``transversity'' distributions $\delta f$ remain virtually
unknown. They are defined \cite{ref:jaffeji,ref:ralston,ref:artru,ref:ratcliffe}
as the differences of probabilities for finding a parton of flavor $f$ at scale
$\mu$ and light-cone momentum fraction $x$ with its spin aligned
($\uparrow\uparrow$) or anti-aligned ($\downarrow\uparrow$)
with that of the transversely polarized nucleon:
\begin{equation}
\label{eq:pdf}
\delta f(x,\mu) \equiv f_{\uparrow\uparrow}(x,\mu) -
                       f_{\downarrow\uparrow}(x,\mu) \; .
\end{equation}
A program of polarized $pp$ collisions is now underway  at the BNL
Relativistic Heavy Ion Collider (RHIC) \cite{ref:rhic}, aiming at
further unraveling the spin structure of the proton. Collisions of
transversely polarized protons are hoped to give information on
transversity through, e.g., the measurement of double-spin
asymmetries
\begin{equation}
\label{eq:att}
A_{\mathrm{TT}} = \frac{\frac{1}{2}
              \left[d\sigma(\uparrow\uparrow) -
              d\sigma(\uparrow\downarrow)\right]}
                                  {\frac{1}{2}
                \left[d\sigma(\uparrow\uparrow) +
               d\sigma(\uparrow\downarrow)\right]}
\equiv \frac{d\delta\sigma}{d\sigma}
\end{equation}
for various reactions with observed produced high-transverse
momentum ($p_T$) or invariant mass. The best-studied, and perhaps
most promising among these, is the Drell-Yan process
\cite{ref:ralston,ref:drellyan,ref:drellyan2}, which offers the largest spin
asymmetries but whose main drawback is the rather moderate event
rate. Other reactions, such as high-$p_T$ prompt-photon, pion, or
jet production, are much more copious, but suffer from fairly
small spin asymmetries
\cite{ref:artru,ref:jaffesaito,ref:ji92,ref:attlo,ref:photonnlo},
due to large contributions from gluon-gluon and quark-gluon
scattering only present in the unpolarized cross section in the
denominator of Eq.~(\ref{eq:att}).

Very recently, it has also been proposed to extract transversity
from measurements of $A_{\mathrm{TT}}$ in transversely polarized
${\bar p}p$ collisions at the planned GSI-FAIR
facility~\cite{ref:pax,ref:assia,ref:anselmino} near Darmstadt,
Germany. For later stages of operations, there are plans to have
an asymmetric $\bar{p}p$ collider, with moderate proton and
antiproton energies of 3.5 and 15~GeV, respectively. So far,
theoretical work has focused on the Drell-Yan process. It was
found that the expected spin asymmetries could be very large,
possibly reaching several tens of 
per cents \cite{ref:anselmino,ref:radici,ref:resum}.
This can be readily understood, because for the GSI kinematics
only partons with rather large momentum fractions scatter off each
other, and in $\bar{p}p$ collisions the relevant lowest order (LO)
process, $q\bar{q}$ annihilation, will receive large contributions
from valence quarks, which are expected to carry strong
polarization. This appears to make the proposed measurements at
GSI particularly interesting for learning about transversity.

The theoretical framework for GSI kinematics is somewhat more
involved than for RHIC, since perturbative all-order resummations
of large logarithmic contributions to the partonic cross sections
are particularly important. For the Drell-Yan process, these have
been addressed in detail recently in~\cite{ref:resum}. In any
case, information from RHIC and from GSI experiments will be
complementary, due to the very different kinematics accessed, and
very likely both will be needed to gain sufficient knowledge about
the $\delta f$ over a large range in $x$.

In this paper, we perform a detailed study of high-$p_T$
single-inclusive pion production in transversely polarized $pp$
and $\bar{p}p$ collisions. In particular, we derive the
next-to-leading order (NLO) QCD corrections to the relevant
partonic cross sections. In general, these are indispensable for
arriving at a firmer theoretical framework for analyzing
experimental data in terms of parton densities. For the
calculation we employ a recently developed ``projection
technique'' for treating the phase space integrals in the presence
of the $\cos(2\Phi)$ azimuthal-angular dependence associated with
transverse polarization \cite{ref:photonnlo}.

We will apply our analytical results in phenomenological studies
for $pp$ and $\bar{p}p$ scattering at RHIC and GSI energies,
respectively. Regarding the latter, to our knowledge our study is
the first to propose accessing transversity via the process
$\bar{p}p \to \pi X$. We hope that such measurements would be
possible with the proposed PAX~\cite{ref:pax} and
ASSIA~\cite{ref:assia} experiments, but this reaction should also
be of great interest for the PANDA Collaboration \cite{ref:panda}.
As we shall see, the spin asymmetry for $\bar{p}p\to\pi X$ is
expected to be much smaller than that for the Drell-Yan process, a
drawback that may be compensated for by the much higher event rates
and, therefore, the much better statistical accuracy. We will also
find, however, that the size of the NLO corrections and the scale
dependence are very significant at GSI energies, so that further
theoretical work will be needed before one can be confident that
high-$p_T$ pion production may be a useful tool to learn about
transversity. If so, combined information from Drell-Yan and from
$\bar{p} p \to \pi X$ (or other produced hadrons) could be useful
since the processes probe different combinations of the
transversity densities.

In the next section we will very briefly review the necessary
technical framework for the computation of the NLO QCD
corrections. Details on the projection technique, which is a
crucial tool for the calculation, can be found in
Ref.~\cite{ref:photonnlo} where it was applied to prompt-photon
production. In Sec.~III we will present phenomenological results
for RHIC and GSI energies. We conclude in Sec.\ IV.

\section{Technical Framework}
%
According to the factorization theorem \cite{ref:fact}, the fully
differential, transverse-spin dependent, single-inclusive cross
section for the reaction $AB \rightarrow \pi X$
for the production of a pion (or any other hadron) with transverse momentum
$p_T$, azimuthal angle $\Phi$ with respect to the initial spin axis,
and pseudorapidity $\eta$ reads at NLO accuracy
\begin{widetext}
\begin{eqnarray}
\label{eq:xsec}
\frac{d^3 \delta \sigma}{dp_T\, d\eta\, d\Phi}
&=& \frac{p_T}{\pi S} \sum_{ab\to cX}\int_{1-V+VW}^1 {d z_c\over z_c^2}
\int_{VW/z_c}^{1-{(1-V)/z_c}}\frac{dv}{v(1-v)}
\int^1_{VW/{v z_c}}\frac{dw}{w}\,\delta f_a(x_a,\mu) \delta f_b(x_b,\mu)
D^\pi_c (z_c, \mu)
\nonumber \\[3mm]
&\times&  \,\left[
\frac{d\delta \hat{\sigma}^{(0)}_{ab\to c X}(v)}{dvd\Phi}
\delta (1-w) + \frac{\alpha_s(\mu)}{\pi} \, \frac{d\delta
\hat{\sigma}^{(1)}_{ab\to c X}(s,v,w,\mu)}{dvdwd\Phi}
\right] \;\; ,
\end{eqnarray}
\end{widetext}
where the sum is over all contributing partonic channels
$ab\to c X$, $AB=pp$ or $\bar{p}p$, 
and with hadron-level variables
\begin{eqnarray}
\nonumber
V&\equiv& 1+\frac{T}{S} \; \;,\;\; W\equiv \frac{-U}{S+T} \;\;,\;\;
S\equiv (P_A+P_B)^2 \;\;,\\[3mm]
T&\equiv& (P_A-P_{\pi})^2 \;\;,\;\; U\equiv (P_B-P_{\pi})^2 \;\; ,
\end{eqnarray}
in obvious notation of the momenta. The corresponding partonic quantities
are given by
\begin{eqnarray} \label{partvar}
\nonumber v&\equiv& 1+\frac{t}{s} \;\;,\;\; w\equiv \frac{-u}{s+t}
\;\;,\;\; s\equiv (p_a+p_b)^2 \;\;,\\[3mm]
t&\equiv& (p_a-p_c)^2 \;\;, \;\; u\equiv (p_b-p_c)^2 \;\;.
\end{eqnarray}
Neglecting all masses, one has the relations
\begin{eqnarray} \label{further}
\nonumber s&=&x_a x_b S \;\;,\;\; t=\frac{x_a}{z_c} T \;\;,\;\;
u=\frac{x_b}{z_c} U \;\;,\\[3mm]
x_a &=& \frac{VW}{vw z_c} \;\;,\;\; x_b = \frac{1-V}{(1-v) z_c}
\;\;.
\end{eqnarray}

The transversity densities in Eq.~(\ref{eq:xsec}) always refer to those
for a parent proton even for $AB=\bar{p}p$, i.e., we use the charge conjugation
property $\delta f_a^{\bar{p}}=\delta f_{\bar{a}}^p$.
The fact that we are observing a specific hadron in the reaction requires
the introduction of additional long-distance functions in Eq.~(\ref{eq:xsec}), 
the parton-to-pion fragmentation functions $D^\pi_c$.
The $d\delta\hat{\sigma}^{(i)}_{ab\to cX}$ are the LO ($i=0$) and NLO
($i=1$) contributions in the cross sections for the partonic reactions
$a b\rightarrow c X$. Finally, $\mu$ collectively denotes the renormalization and
factorization scales, which we will always take as equal for simplicity.

We now give a few technical details of the NLO calculation.
Projection on a definite polarization state for the initial
partons involves the Dirac matrix $\gamma_5$. It is well known that
in dimensional regularization, which we will use to regularize
the ultraviolet, infrared, and collinear singularities at intermediate
stages of the calculation, the treatment of $\gamma_5$ is in general
a subtle issue. However, owing to the chirally odd nature of transversity,
in our calculation all Dirac traces contain {\it two}
$\gamma_5$ matrices, and, therefore, using the ``HVBM scheme''~\cite{ref:hvbm}
or a naive, totally anticommuting $\gamma_5$ in $n\neq 4$ dimensions 
must give the same results which is also a useful check for the correctness
of the calculation.

The transverse polarization vectors of the initial hadrons
give rise to a characteristic dependence of the cross section on the
azimuthal angle $\Phi$ of the observed particle.
In the hadronic center-of-mass system (c.m.s.) frame, taking the initial hadrons
along the $\pm z$ axis and their spin vectors in $\pm x$ direction,
the $\Phi$-dependence is of the form $\cos (2\Phi)$.
Integration over $\Phi$ is therefore
not appropriate. Keeping $\Phi$ fixed in the NLO calculation is however
very cumbersome since standard techniques developed in the literature for
performing NLO phase-space integrations rely on the choice of
particular reference frames different from the one specified above.
In \cite{ref:photonnlo} we developed a general projection method that involves
integration over all $\Phi$, thereby allowing to keep the benefits of the
standard phase space integration techniques.  The trick, and the 
virtue of our method, is to project out the dependence of the matrix elements 
on the spin vectors in a {\em covariant} way, by multiplying with a covariant 
expression for the $\cos(2 \Phi)$ term, and to then carry out the complete 
phase space integrals.

To be more specific, we note that because of the $\cos(2\Phi)$
dependence we have the identity
\begin{equation}
\label{eq2}
\frac{d^3\delta \sigma}{dp_T d\eta d\Phi}\;\equiv\;
\cos (2\Phi)
\int_0^{2\pi}\,d\Phi^\prime \,\frac{\cos (2\Phi^\prime)}{\pi}\;
\frac{d^3\delta \sigma}{dp_T d\eta d\Phi^\prime}
\; .
\end{equation}
The $\cos (2\Phi)$ dependence actually arises through the covariant expression
\begin{equation}
\label{eq5}
{\cal F}(p_c,s_a,s_b)\;=\;
\frac{s}{t u}
\,\left[ 2 \,(p_c\cdot s_a)\, (p_c\cdot s_b)\; +\;
\frac{t u}{s} \,(s_a \cdot s_b) \right] \;,
\end{equation}
where the $s_i$ ($i=a,b$) are the initial transverse spin vectors which 
satisfy $s_i\cdot p_a= s_i\cdot p_b=0$ and $s_a^2=s_b^2=-1$.
${\cal F}(p_c,s_a,s_b)$ reduces to $\cos (2\Phi)$ in the hadronic c.m.s.\ frame.
We may, therefore, use ${\cal F}(p_c,s_a,s_b)/\pi$ instead of the explicit
$\cos (2\Phi)/\pi$ in the ``projector'' in the integrand of Eq.~(\ref{eq2}).
For any contributing partonic channel we multiply the squared matrix element
for transversely polarized initial partons, $\delta |M|_{ab\to cX}^2$,
by ${\cal F} (p_c,s_a,s_b)/\pi$.  The resulting expression may then
be integrated over the full azimuthal phase space in a covariant way 
without producing a vanishing result, unlike the case of $\delta |M|^2$ itself;
see Ref.~{\cite{ref:photonnlo} for further details.
It is crucial here that the other observed (``fixed'') quantities, 
the hadron's transverse momentum $p_T$ and 
pseudorapidity $\eta$, are determined entirely by scalar products $(p_a\cdot
p_{c})$ and $(p_b\cdot p_{c})$, independently of the spin vectors $s_{a,b}$.

Our method becomes particularly convenient for treating the 
$2\to 3$ scattering contributions arising at NLO where one has an
additional phase space integral over the second unobserved parton in the
final-state. After applying the projection method 
we can perform all phase space integrations by employing techniques
familiar from the corresponding calculations in the unpolarized
and longitudinally polarized cases~\cite{ref:nlostuff,ref:nlopion}.
We note that as a non-trivial check on our calculation
we have also integrated all squared matrix elements over the
spin vectors {\em without} using any projector at all. This
amounts to integrating $\cos(2\Phi)$ over all $0\leq\Phi\leq 2\pi$,
and, as expected, the final answer is zero.

The use of dimensional regularization is
straightforward in all this.
Ultraviolet poles in the virtual diagrams are removed by the
renormalization of the strong coupling constant.
Infrared singularities cancel in the sum between virtual
and real-emission diagrams. After this cancellation,
only collinear poles are left. From the factorization theorem it follows that
these need to be factored into the parton distribution and fragmentation
functions.  This is a standard procedure which we have also described in quite some
detail in ~\cite{ref:nlopion,ref:photonnlo}. We use the $\overline{\rm{MS}}$ scheme
throughout.

After factorization, we arrive at the final result, the finite partonic
NLO hard scattering cross sections. There are all in all five subprocesses that
contribute for transverse polarization:
\begin{eqnarray}
qq &\to& qX, \nonumber \\
q{\bar q} &\to& qX,\nonumber\\
q{\bar q} &\to& q'X,\nonumber\\
q{\bar q} &\to& gX ,\nonumber\\
qq &\to& gX \; ,
\end{eqnarray}
where at NLO in each case $X$ denotes a one- or two-parton final state,
summed over all possibilities and integrated over its phase space.
The first four of these reactions are present at LO already. The corresponding
LO transversity cross sections may be found
in \cite{ref:jaffesaito,ref:ji92,ref:attlo,ref:attold}.
The last subprocess appears for the first time at NLO. For each of the five
subprocesses, the NLO expression for the transversely polarized cross section
can be cast into the following form:
\begin{widetext}
\begin{eqnarray}
&&  s\, \frac{d \delta \hat{\sigma}_{ab\to cX}^{(1)}
(s,v,w,\mu)}{dvdwd\Phi}\,=\,\cos(2\Phi)\,
{\left(\frac{\alpha_s(\mu)}{\pi}\right )}^2
  \Bigg[ \left(  A_0  \delta (1-w) +   B_0
\frac{1}{(1-w)_+} +  C_0 \right) \ln \frac{\mu^2}{s}
\nonumber \\[3mm]
&&+A \delta (1-w) +B \frac{1}{(1-w)_+}+  C
+ D\left(
\frac{\ln (1-w)}{1-w} \right)_+ + E \ln w + F \ln v
+ G \ln (1-v)\nonumber \\[3mm]
&&+ H \ln(1-w)  + I  \ln (1-vw) +J  \ln (1-v+vw) +K \frac{\ln
w}{1-w} + L \frac{\ln \frac{1-v}{1-vw}}{1-w} +M \frac{\ln
(1-v+vw)}{1-w}  \Bigg] \:\:\: , \label{final}
\end{eqnarray}
\end{widetext}
where the ``plus''-distribution is defined in the usual way over
the interval $[0,1]$. All coefficients in Eq.~(\ref{final})
are functions of $v$ and $w$, except those
multiplying the distributions $\delta(1-w)$, $1/(1-w)_+$,
$\left[ \ln(1-w)/(1-w)\right]_+$ which may be written as
functions just of $v$. Terms with distributions are present only
for the subprocesses that already contribute at the Born level.
The coefficients are available upon request as a {\sc Fortran} code
from the authors.

\section{Phenomenological Results}

We now present some phenomenological results for
single-inclusive pion production in transversely polarized
$pp$ collisions at RHIC ($\sqrt{S}=200$ and 500~GeV) and 
asymmetric $\bar{p}p$ collisions at the planned GSI-FAIR facility with 
proton and antiproton energies of $3.5$ GeV and 15~GeV, respectively.

Since nothing is known experimentally about transversity so far,
we need to model the $\delta f$ for our study. Guidance is provided by the Soffer
inequality \cite{ref:soffer}
\begin{equation}
\label{eq:soffer}
2\left|\delta q(x)\right| \leq q(x) + \Delta q(x)
\end{equation}
which gives bounds for each $\delta f$.
As in \cite{ref:attlo,ref:photonnlo} we utilize this inequality by saturating
the bound at some low input scale $\mu_0\simeq 0.6\,\mathrm{GeV}$, choosing
all signs to be positive, and using
the NLO (LO) GRV \cite{ref:grv} and GRSV (``standard scenario'')
\cite{ref:grsv} densities $q(x,\mu_0)$ and $\Delta q(x,\mu_0)$,
respectively. For $\mu>\mu_0$ the transversity densities $\delta f(x,\mu)$
are then obtained by evolving them at LO or NLO. We refer
the reader to \cite{ref:drellyan2,ref:attlo} for more details on our
model distributions. We note that we will always perform the
NLO (LO) calculations using NLO (LO) parton distribution functions
and the two-loop (one-loop) expression for $\alpha_s$.
We use the pion fragmentation functions of Ref.~\cite{ref:frag} which
has both a LO and an NLO set. They provide a very good description of
the recent RHIC data on unpolarized neutral-pion production \cite{ref:rhicdata}.

Figure~\ref{fig:rhicxsec} shows our estimates for
the transversely polarized single-inclusive pion production cross sections at
LO and NLO for the two different c.m.s.\ energies at RHIC. We have
integrated over the range $|\eta|\le 0.38$ in pseudorapidity, appropriate
for measurements  with the PHENIX detector. Since only half of the pion's 
azimuthal angle is covered, we integrate
over the two quadrants $-\pi/4<\Phi<\pi/4$ and $3\pi/4<\Phi<5\pi/4$, which
gives $\left(\int_{-\pi/4}^{\pi/4}+
\int_{3\pi/4}^{5\pi/4} \right) \cos(2\Phi) d\Phi=2$.
We have also varied simultaneously the factorization/renormalization scales $\mu$
in Eq.~(\ref{eq:xsec}) within $p_T \le \mu \le 4p_T$; a significant decrease
of scale dependence is observed when going from LO to NLO.
\begin{figure}[t!]
\centering
\includegraphics[width=8.5cm]{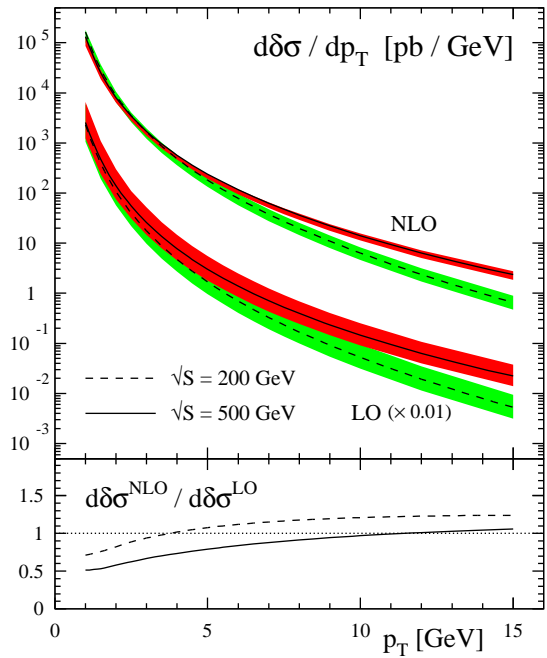}
\caption{\label{fig:rhicxsec}
\sf Transversely polarized single-inclusive neutral-pion
production cross sections in LO and NLO for $\sqrt{S}=200$
and 500 GeV collisions at RHIC. The LO results have been 
scaled by a factor of 0.01.
The shaded bands represent the changes if the
scale $\mu$ is varied in the range $p_T \le \mu \le 4p_T$.
The lower panel shows the ratios of the NLO and LO results 
for both c.m.s.\ energies.}
\end{figure}

The lower part of the Figure~\ref{fig:rhicxsec} displays the so-called ``$K$-factor'',
defined as usual as the ratio of the NLO to the LO cross section,
for the scale choice $\mu=2 p_T$.
Except for small $p_T$, where the NLO corrections lead to a significant reduction of 
the cross section, the $K$-factor turns out to be rather moderate and close to unity.
It is known that the $K$-factor for the unpolarized
cross section is significantly larger than one at RHIC energies, see, e.g., Fig.\ 4 in 
Ref.~\cite{ref:nlopion} for $\sqrt{S}=200\,\mathrm{GeV}$, mostly because
of large corrections found for gluon-initiated partonic channels. 
Therefore, one expects that the double-spin asymmetry $A_{\mathrm{TT}}$ at RHIC 
will decrease when going from LO to NLO. Indeed, as Fig.~\ref{fig:attrhic} shows, this is the
case. Here we used the CTEQ6M (CTEQ6L1)~\cite{ref:cteq6} set of unpolarized parton
distributions to calculate the corresponding NLO (LO) unpolarized cross section.
We have chosen the scale $\mu=p_T$ which leads to the largest cross sections in
Fig.~\ref{fig:rhicxsec}. 
\begin{figure}[t!]
\centering
\includegraphics[width=8cm]{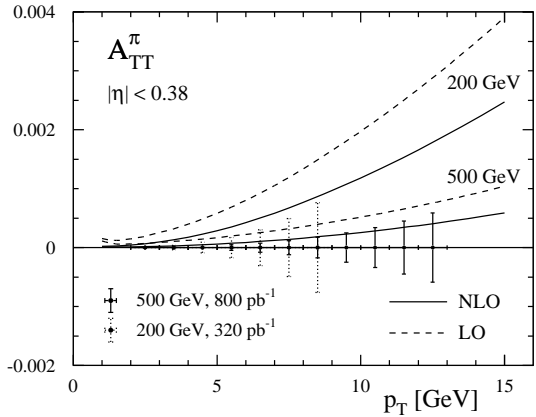}
\caption{\label{fig:attrhic}
\sf Upper bounds for the double-transverse spin asymmetry $A_{\mathrm{TT}}$
corresponding to Fig.~\ref{fig:rhicxsec}. The ``error bars'' indicate the statistical
accuracy that might be achievable at RHIC (see text).}
\end{figure}
We also indicate in Fig.~\ref{fig:attrhic} an estimate of the statistical accuracy
that might be achievable at RHIC, based on
\begin{equation} \label{eq:error}
\delta A_{\mathrm{TT}}
\simeq \frac{1}{P_A P_B \sqrt{{\cal L}\,\sigma_{\rm bin}}} \; ,
\end{equation}
with beam polarizations $P_{A,B}$ of $70\%$, and an integrated luminosity ${\cal L}$ 
of 320 and $800\,\mathrm{pb}^{-1}$ for c.m.s.\ energies of
$\sqrt{S}=200$ and 500~GeV, respectively. 
$\sigma_{\rm bin}$ denotes the unpolarized cross section integrated over the 
$p_T$-bin for which the error is to be determined. 
Clearly, the statistics would be sufficient
to measure even asymmetries as small as the ones shown in Fig.~\ref{fig:attrhic}.
However, it is likely that the systematic error on spin asymmetries
at RHIC will not be much smaller than $10^{-3}$, in which case it would
appear to be very difficult to access transversity from $A_{\mathrm{TT}}$
for single-inclusive pion production.
We stress again that the results shown in Fig.~\ref{fig:attrhic} are upper bounds,
at least within the GRV/GRSV framework with its low input scale for the
evolution. If the bound in Eq.~(\ref{eq:soffer}) turns out to be not saturated 
at that scale the asymmetries would be even smaller. On the other hand, if we
used transversity densities that saturate Eq.~(\ref{eq:soffer}) at a higher scale,
say $\mu_0\simeq 1\,\mathrm{GeV}$, the results for $A_{\mathrm{TT}}$ would be 
somewhat larger. In any case the measurement is very challenging at RHIC.

\begin{figure*}[hbt!]
\centering
\includegraphics[width=6.7cm,clip]{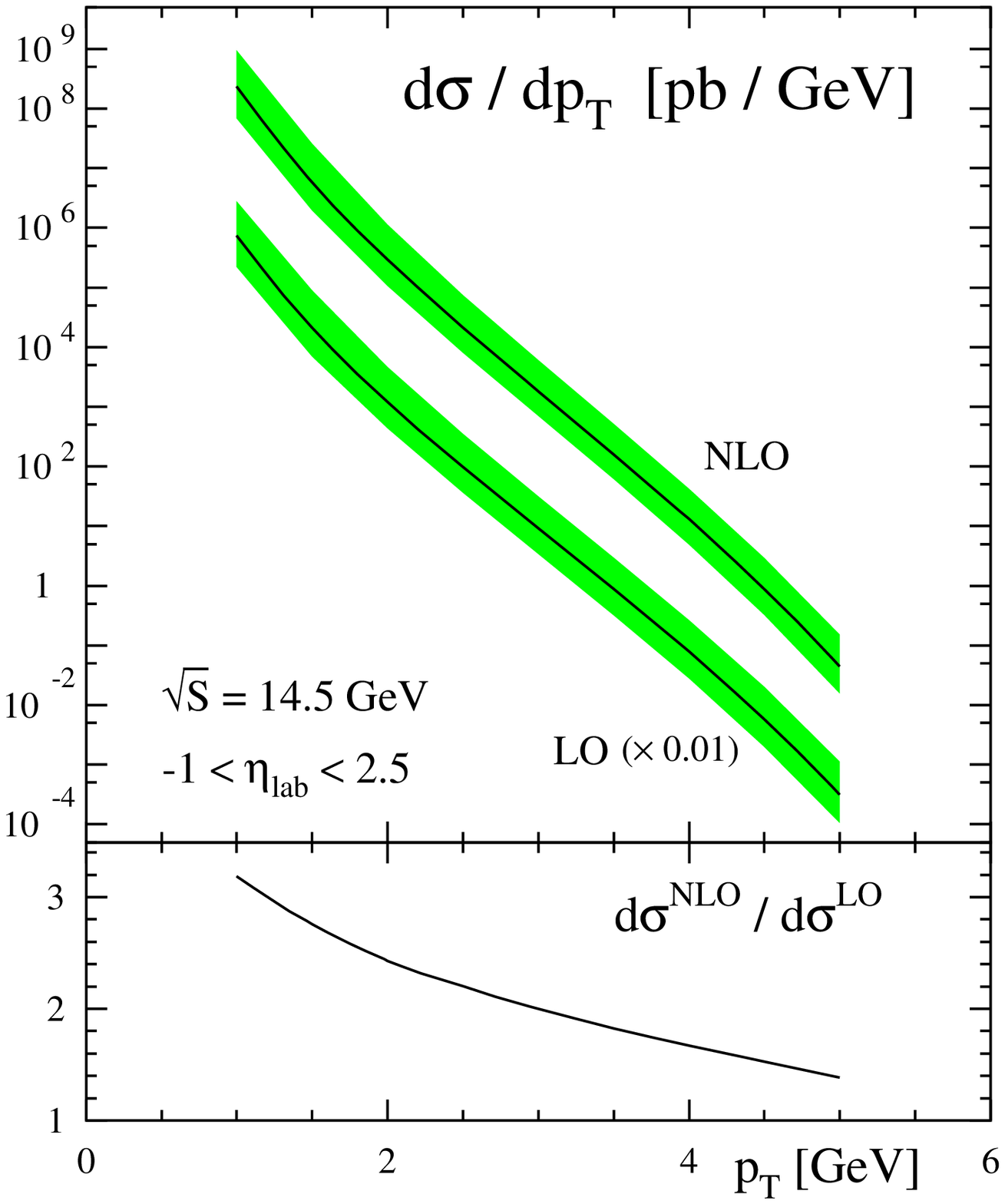}
\hspace{0.2cm}
\includegraphics[width=6.7cm,clip]{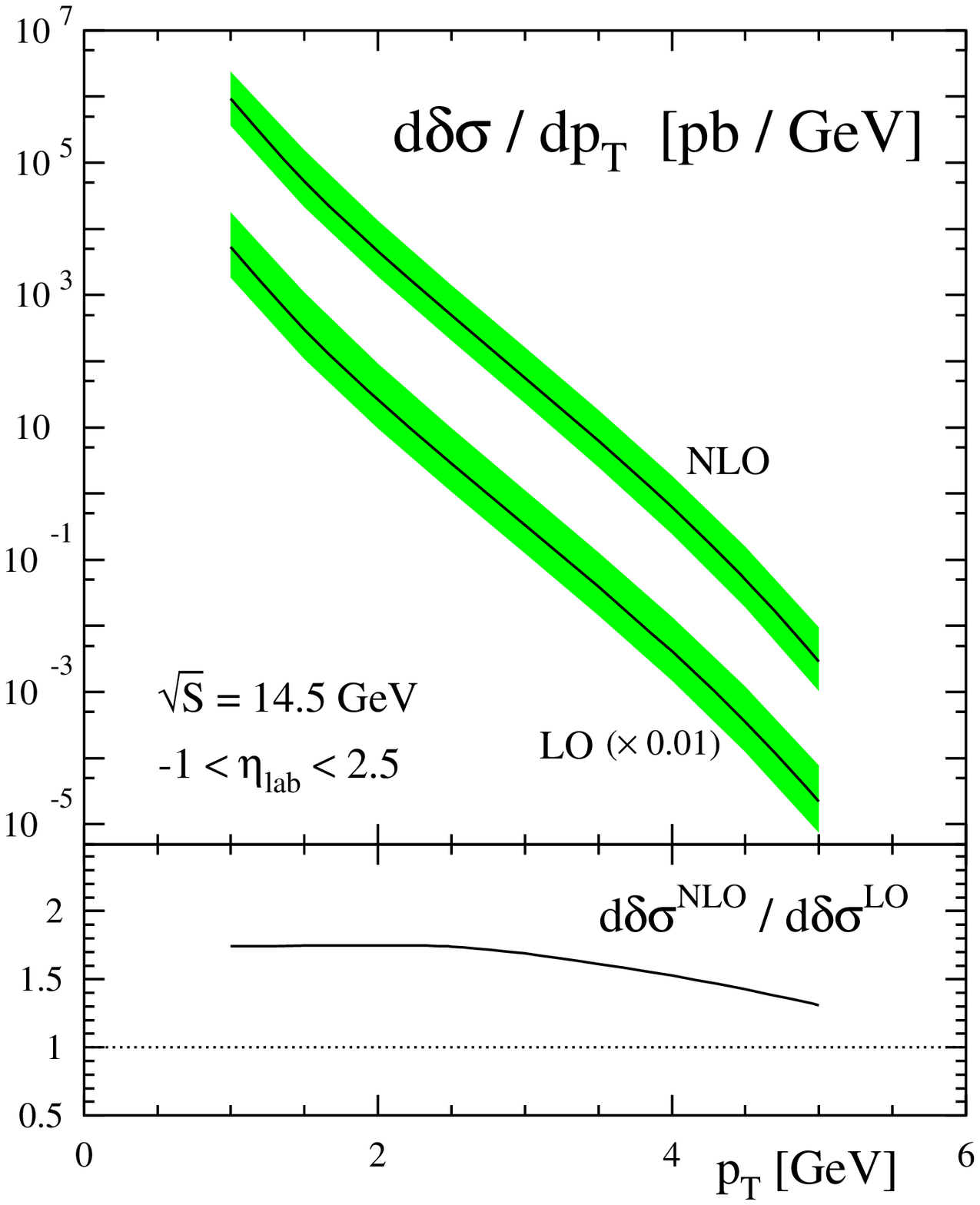}
\vspace{-0.4cm}
\caption{\label{fig:xsecgsi}
\sf Unpolarized (left) and transversely polarized
(right) single-inclusive neutral-pion production cross sections at LO and NLO at
the GSI. The LO results have been scaled by a factor of 0.01.
The shaded bands represent the theoretical uncertainty if the
factorization/renormalization scale
$\mu$ is varied in the range $p_T \le \mu \le 4p_T$.
The lower panel shows the ratios of the NLO and LO results in each case, 
using $\mu=2 p_T$.}
\end{figure*}
We now turn to transversely polarized $\bar{p}p$ collisions with $\sqrt{S}=
14.5$~GeV at the planned GSI-FAIR facility. We first note that for this rather moderate
c.m.s.\ energy the pion transverse momentum can at most reach $7.25$~GeV,
at mid rapidity. In our study we integrate over 
$-1 < {\eta}_{{\mathrm{lab}}} < 2.5$, where ${\eta}_{{\mathrm{lab}}}$ is the 
pseudorapidity of the pion in the laboratory frame. We count positive rapidity 
in the forward direction of the antiproton. For the asymmetric collider option 
we consider here, ${\eta}_{{\mathrm{lab}}}$ is related to the c.m.s.\ pseudorapidity
$\eta$ via
\begin{equation}
{\eta}_{{\mathrm{lab}}}=\eta+\frac{1}{2}\ln\frac{E_{\bar{p}}}{E_p}\; ,
\end{equation}
where $E_{\bar{p}}$, $E_p$ are the antiproton and proton energies. The
rapidity interval we use is roughly symmetric in c.m.s.\ pseudorapidities,
$|\eta|\lesssim 1.75$.

Figure~\ref{fig:xsecgsi} shows our results for the unpolarized (left) and
transversely polarized (right) cross sections at NLO and LO, as functions
of $p_T$. For the calculation in the unpolarized case we have chosen
the GRV~\cite{ref:grv} parton distributions. This choice is motivated by
our ansatz for the transversity distributions, for which we had also used
the GRV densities when saturating the Soffer inequality, Eq.~(\ref{eq:soffer}).
Unlike at RHIC energies, at $\sqrt{S}=14.5$~GeV and $p_T$ of several GeV, rather large 
momentum fractions $x_{a,b}$ of the partons are probed in Eq.~(\ref{eq:xsec}), where 
the polarized and unpolarized parton densities for a given parton type are expected to become 
similar~\cite{ref:largex}. It then appears most sensible to use the same parton 
distributions in the unpolarized case that we used when modeling the transversity densities.
In this way we avoid any artificial effects in the NLO corrections and 
$A_{\mathrm{TT}}$ induced by a mismatch in the $x\to 1$ behavior of the parton densities used
in the calculation.

The shaded bands in the upper panels of Fig.~\ref{fig:xsecgsi} again
indicate the uncertainties due to scale variation
in the range $p_T \leq \mu \leq 4 p_T$. One can see that for both, the
unpolarized and the polarized cross sections, the scale dependence does not
really improve from LO to NLO. This is a characteristic feature in low-order
perturbative calculations of cross sections for lower fixed-target energies,
suggesting that corrections beyond NLO are still very significant. Indeed, it
was recently shown~\cite{ref:resumpion} that for inclusive-hadron production in the
fixed-target regime certain double-logarithmic corrections to the partonic cross
sections are important at each order of perturbation theory, and need
to be resummed to all orders to achieve an adequate theoretical description.
Such a resummation will be required in particular in the case we are considering
here and would be very desirable for the future, along with a study of
power corrections.  We emphasize that when the proposed measurements of 
$A_{\mathrm{TT}}$ will be performed, it will be crucial to have precise 
measurements also of the unpolarized
cross section, in order to test the theoretical framework. Only if the
theory is sufficiently understood will data on $A_{\mathrm{TT}}$ become useful
for determining transversity. Similar comparisons of
data~\cite{ref:rhicdata} and theoretical calculations for the unpolarized
neutral-pion cross section at RHIC have shown an excellent agreement even down to
fairly low pion transverse momenta, which has indeed provided much confidence
that the calculations based on partonic hard-scattering are adequate, so that
spin asymmetries measured at RHIC determine the spin-dependent parton distributions of
the proton.

\begin{figure}
\centering
\includegraphics[width=8.5cm,clip]{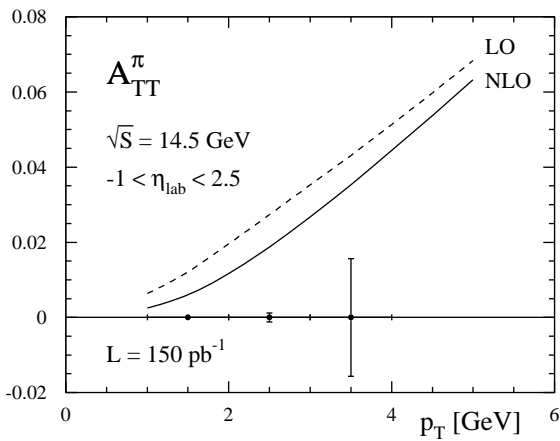}
\vspace{-0.4cm}
\caption{\sf Upper bounds for the transverse double-spin asymmetry
$A_{\mathrm{TT}}$ for single-inclusive neutral-pion production in LO and NLO 
at GSI-FAIR. The ``error bars'' indicate the expected statistical accuracy for 
bins in $p_T$ (see text).
\label{att1}}
\end{figure}
The lower parts of Fig.~\ref{fig:xsecgsi} display the corresponding
$K$-factors at scale $\mu=2 p_T$. We note that these decrease as $p_T$ increases,
which is related entirely to the different behavior of the LO and NLO parton
distributions at large $x$. Had we chosen the same parton distributions
at LO and NLO, the $K$-factors would actually slightly {\it increase}
with $p_T$, as a result of the large double-logarithmic corrections mentioned
above, which first arise at NLO and are known to enhance the cross
section.

Figure~\ref{att1} shows upper bounds--again within the GRV/GRSV framework--for 
the double-spin asymmetry $A_{\mathrm{TT}}$ 
for the GSI-FAIR facility. The scale $\mu$ is again set to $p_T$. We also give expectations
for the statistical errors that may be achievable in experiment. We have calculated 
these using Eq.~(\ref{eq:error}), assuming an integrated luminosity of 
${\cal L}=150\,\mathrm{pb}^{-1}$, and beam polarizations of $30\%$ and $50 \%$ for the 
antiprotons and protons, respectively.

\section{Conclusions}
%

We have presented in this paper the complete NLO QCD corrections for the
partonic hard-scattering cross sections relevant for the
double-spin asymmetry $A_{\mathrm{TT}}$ for single-inclusive high-$p_T$
pion production in collisions of transversely polarized hadrons. This asymmetry
could be a tool to determine the transversity distributions of the nucleon.
Our calculation is based on a largely analytical evaluation
of the NLO partonic cross sections, and we have used a projection technique
for treating the characteristic azimuthal-angle dependence introduced by the transverse spin vectors.

In our phenomenological studies we found that the spin asymmetry
$A_{\mathrm{TT}}$ is expected to be very small in $pp$ collisions
at RHIC and even decreases when going from LO to NLO, due to a larger
$K$-factor in the unpolarized case. We have also studied $A_{\mathrm{TT}}$ for
possibly forthcoming transversely polarized $\bar{p}p$ collisions in
an asymmetric collider mode at the GSI-FAIR facility. Here, the spin asymmetry
may be much larger, but it will be crucial in the future to investigate the effects
of all-order resummations of large Sudakov logarithms. Detailed measurements of the
unpolarized cross sections will be essential for testing the 
applicability of the theoretical framework at the moderate c.m.s.\ energies available
at GSI-FAIR.

\section*{Acknowledgments}
%
We thank F.\ Rathmann, E.\ Reya, and W.\ van Neerven for useful
discussions. W.V.\ is grateful to RIKEN, Brookhaven National
Laboratory and the U.S.\ Department of Energy (contract number
DE-AC02-98CH10886) for providing the facilities essential for the
completion of this work. This work was supported in part 
by the ``Bundesministerium f\"{u}r Bildung und Forschung (BMBF)''
and the ``Foundation for Fundamental Research on Matter'' (FOM), The
Netherlands.


%
\newpage


\begin{thebibliography}{99}
%
\bibitem{ref:jaffeji} R.L.\ Jaffe and X.\ Ji,
Phys. Rev. Lett. {\bf 67}, 552 (1991);
Nucl. Phys. {\bf B375}, 527 (1992).
%
\bibitem{ref:ralston} J.P.\ Ralston and D.E.\ Soper,
Nucl. Phys. {\bf B152}, 109 (1979).
%
\bibitem{ref:artru} X.\ Artru and M.\ Mekhfi, Z. Phys. {\bf C45}, 669 (1990).
%
\bibitem{ref:ratcliffe} A comprehensive review on transversity can be
found in: V.\ Barone, A.\ Drago, and P.G.\ Ratcliffe,
Phys. Rept. {\bf 359}, 1 (2002).
%
\bibitem{ref:rhic} See, for example: G.\ Bunce, N.\ Saito, J.\ Soffer, and
W.\ Vogelsang, Annu.\ Rev.\ Nucl.\ Part.\ Sci.\ {\bf 50}, 525 (2000);
C. Aidala  {\em et al.}, {\it Research Plan for Spin Physics at RHIC},
{\tt http://spin.riken.bnl.gov/rsc/report/masterspin.pdf}
%
\bibitem{ref:drellyan} J.L.~Cortes, B.~Pire, and J.P.~Ralston,
Z.\ Phys.\  {\bf C55}, 409 (1992); W.~Vogelsang and A.~Weber,
Phys.\ Rev.\  {\bf D48}, 2073 (1993); A.P.~Contogouris, B.~Kamal,
and Z.~Merebashvili, Phys.\ Lett.\  {\bf B337}, 169 (1994);
V.~Barone, T.~Calarco, and A.~Drago, Phys.\ Rev.\  {\bf D56}, 527
(1997).
%
\bibitem{ref:drellyan2} O.~Martin, A.~Sch\"{a}fer, M.~Stratmann, and W.~Vogelsang,
Phys.\ Rev.\ {\bf D57}, 3084 (1998); {\bf D60}, 117502 (1999).
%
\bibitem{ref:jaffesaito} R.L.\ Jaffe and N.\ Saito, Phys. Lett. {\bf B382}, 165 (1996).
%
\bibitem{ref:ji92} X.\ Ji, Phys. Lett. {\bf B284}, 137 (1992).
%
\bibitem{ref:attlo} J.\ Soffer, M.\ Stratmann, and W.\ Vogelsang,
Phys. Rev. {\bf D65}, 114024 (2002).
%
\bibitem{ref:photonnlo} A. Mukherjee, M. Stratmann, and W. Vogelsang, Phys. Rev.
{\bf D67}, 114006 (2003).
%
\bibitem{ref:pax} GSI-PAX Collaboration,
P.~Lenisa and F.~Rathmann (spokespersons) {\em et al.}, Technical
Proposal, {\tt hep-ex/0505054}; see also: F.~Rathmann and
P.~Lenisa, in {\em Proceedings of the 16th International Spin
Physics Symposium (SPIN 2004)}, Trieste, Italy, 2004, {\tt
hep-ex/0412078}; P.~Lenisa {\it et al.}, in {\em Proceedings of
the 2nd High-Energy Physics Conference in Madagascar (HEP-MAD
04)}, Antananarivo, Madagascar, 2004, eConf {\bf C0409272}, 014
(2004).
%
\bibitem{ref:assia} GSI-ASSIA Collaboration,
R.\ Bertini (spokes\-person) {\em et al.}, Technical Proposal, {\tt
http://www.gsi.de/documents/\newline DOC-2004-Jan-152-1.ps}; GSI-ASSIA
Collaboration, M.\ Maggiora, in {\em Proceedings of the Conference
on Spin and Symmetry}, Prague, 2004, {\tt hep-ex/0504011}.
%
\bibitem{ref:anselmino}
M.~Anselmino, V.~Barone, A.~Drago, and N.N.~Nikolaev, Phys.\
Lett.\  {\bf B594}, 97 (2004); A.V.~Efremov, K.~Goeke, and
P.~Schweitzer, Eur.\ Phys.\ J.\ {\bf C35}, 207 (2004).
%
\bibitem{ref:radici} P.G.~Ratcliffe, Eur.\ Phys.\ J.\ {\bf C41}, 319 (2005);
A.~Bianconi and M.~Radici, {\tt hep-ph/0504261}.
%
\bibitem{ref:resum} H. Shimizu, G. Sterman, W. Vogelsang, and H. Yokoya,
Phys.\ Rev.\ {\bf D71}, 114007 (2005).
%
\bibitem{ref:panda} GSI-PANDA Collaboration, U.\ Wiedner
(spokes\-person), Technical Progress Report, {\tt
http://www.gsi.de/fair/\newline experiments/hesr-panda/index.html}.
%
\bibitem{ref:fact} J.C.\ Collins, D.E.\ Soper, and G.\ Sterman,
Nucl.\ Phys.\  {\bf B261}, 104 (1985); Nucl.\ Phys.\ B {\bf 308},
833 (1988); in {\em Perturbative Quantum Chromodynamics},
A.H.~Mueller (ed.), World Scientific Publ., Singapore, 1989, p.1
[{\tt hep-ph/0409313}]; G.T.~Bodwin, Phys.\ Rev.\ {\bf D31}, 2616
(1985); {\bf D34}, 3932(E) (1986)]; J.C.~Collins, Nucl.\ Phys.\ 
{\bf B394}, 169 (1993).
%
\bibitem{ref:hvbm} G.\ 't Hooft and M. Veltman, Nucl. Phys. {\bf B44},
189 (1972); P. Breitenlohner and D. Maison, Commun. Math. Phys. {\bf 52},
11 (1977).
%
\bibitem{ref:nlostuff} R.K.\ Ellis, M.A.\ Furman, H.E.\ Haber,
and I.\ Hinchliffe, Nucl. Phys. {\bf B173}, 397 (1980);
D.W.\ Duke and J.F.\ Owens, Phys. Rev. {\bf D26}, 1600 (1982);
{\bf D28}, 1227(E) (1983);
P.~Aurenche, A.~Douiri, R.~Baier, M.~Fontannaz, and D.~Schiff,
Phys.\ Lett.\ {\bf B140}, 87 (1984); 
P.~Aurenche, R.~Baier, M.~Fontannaz, and D.~Schiff, Nucl.\ Phys.\
{\bf B297}, 661 (1988);
L.~E.~Gordon and W.~Vogelsang, Phys.\ Rev.\ {\bf D48}, 3136 (1993).
%
\bibitem{ref:nlopion} B.~J\"{a}ger, A.~Sch\"{a}fer,
M.~Stratmann, and W.~Vogelsang, Phys.\ Rev.\ {\bf D67}, 054005
(2003).
%
\bibitem{ref:attold} K.\ Hidaka, E.\ Monsay, and D.\ Sivers,
Phys. Rev. {\bf D19}, 1503 (1979).
%
\bibitem{ref:soffer} J.\ Soffer, Phys. Rev. Lett. {\bf 74}, 1292 (1995);
D.\ Sivers, Phys. Rev. {\bf D51}, 4880 (1995).
%
\bibitem{ref:grv} M.\ Gl\"{u}ck, E.\ Reya, and A.\ Vogt,
Eur.\ Phys.\ J.\ {\bf C5}, 461 (1998).
%
\bibitem{ref:grsv}  M.\ Gl\"{u}ck, E.\ Reya, M.\ Stratmann, and
W.\ Vogelsang, Phys. Rev. {\bf D63}, 094005 (2001).
%
\bibitem{ref:frag} B. A. \ Kniehl, G. \ Kramer, and B. \ P\"otter, Nucl. Phys. {\bf
B582}, 514 (2000).
%
\bibitem{ref:rhicdata} PHENIX Collaboration, S.S.\ Adler {\em et al.},
Phys. Rev. Lett. {\bf 91}, 241803 (2003);
STAR Collaboration, J.\ Adams {\em et al.},
Phys. Rev. Lett. {\bf 92}, 171801 (2004);
STAR Collaboration, G.~Rakness, presented at the {\em XXXX Rencontres de Moriond on
QCD and High Energy Hadronic Interactions}, La Thuile, Italy,
2005, {\tt hep-ex/0505062}.
%
\bibitem{ref:cteq6} J.\ Pumplin {\it et al.}, JHEP 0207, 012 (2002).
%
\bibitem{ref:largex} G.\ Farrar and D.R.\ Jackson, Phys. Rev. Lett. {\bf 35}, 1416
(1975); S.J.\ Brodsky, M.\ Burkardt, and I.\ Schmidt, Nucl.
Phys. {\bf B441}, 197 (1995); see also: E.\ Leader, A.V.\ Sidorov,
and D.B.\ Stamenov, Int. J. Mod. Phys. {\bf A13}, 5573 (1998).
%
\bibitem{ref:resumpion} D.~de Florian and W.~Vogelsang, Phys. Rev. {\bf D71},
114004 (2005).
%
\end{thebibliography}
\end{document}